\documentclass[a4paper,12pt]{article}
\usepackage{amsmath}
\usepackage{graphicx}
\usepackage{rotating}
\usepackage{wasysym}
\DeclareGraphicsExtensions{.pdf,.png,.jpg}
\usepackage{chicago}
\usepackage{klett}
\usepackage{def_the}
\usepackage{Fig_size_A}

\title{Information-Theoretic Model for the\\ \vspace{0.5cm} Growth Optimal Portfolio of Stocks}
\author{Eckhard Platen and Kevin Fergusson}
\begin{document}
	\thispagestyle{empty} \vspace*{1.0cm}
	
	\begin{center}
		{\LARGE\bf Information-Theoretic Model for the\\ \vspace{0.5cm} Growth Optimal Portfolio of Stocks} \\\vspace{0.5cm}  
\end{center}

\vspace*{.5cm}
\begin{center}{\large \renewcommand{\thefootnote}{\arabic{footnote}} {\bf
	 Eckhard Platen\footnote{UTS Business School and School of Mathematical and Physical Sciences,\\ University of Technology, Sydney, Broadway, Ultimo, New South Wales, Australia,\\  and College of Business and Economics, The Australian National University, Canberra, Australia, {eckhard.platen}@uts.edu.au} and Kevin Fergusson\footnote{Centre for Data Analytics, Bond University, Gold Coast, Queensland, Australia,\\  {kferguss}@bond.edu.au}
} }\\
\vspace*{.5cm}
\today
\vspace*{2.5cm}

\end{center}

\begin{minipage}[t]{13cm} Drawing on information-theoretic principles and empirical evidence, a proxy for the growth-optimal portfolio of stocks (the benchmark) is constructed and modeled. The derivative of its activity time captures overreactions in trading behavior and follows a fast-moving, quasi-stationary scalar diffusion. The benchmark's log-returns are predicted to follow a Student $t$ distribution with degrees of freedom slightly less than four, and this prediction is tested empirically. With the benchmark as num\'eraire, the derived model enables benchmark-neutral pricing and hedging.
\end{minipage}
\vspace*{0.5cm}


\vspace*{0.5cm}
\vspace*{0.5cm}

\newpage
\section{Introduction}
For the pricing of contingent claims, it is paramount to have a theoretically sound and, in practice, reliable pricing method. The theoretical assumption of the absence of {\em free lunches with vanishing risk} (FLVRs) in the Fundamental Theorem of Asset Pricing of \citeN{DelbaenSc98} underpins the risk-neutral pricing method. In \citeN{PlatenFe25a}, this crucial assumption has been shown not to be realistic. Risk-neutral long-term derivative prices turn out to be significantly overpriced; see, e.g., \citeN{BaroneadesiPlSa24}, \citeN{Fergusson20}, and \citeN{FergussonPl23}. Since long-term pricing is important in areas like pensions, and risk-neutral pricing is theoretically unsatisfactory for extreme-maturity contingent claims, {\em benchmark-neutral} (BN) pricing has been proposed in \citeN{Platen25a} as  alternative. BN pricing employs as num\'eraire the {\em benchmark}, which is the {\em growth optimal portfolio} (GOP) of the market formed by the stocks.\\

  The GOP of a market is interchangeably called the Kelly portfolio, the expected logarithmic utility-maximizing portfolio, or the num\'eraire portfolio; see, e.g., \citeN{Kelly56}, \citeN{Merton71}, \citeN{Long90}, \citeN{Becherer01}, \citeN{Platen06ba}, \citeN{KaratzasKa07}, \citeN{HulleySc10}, and \citeN{MacLeanThZi11}. It follows from Theorem 10.5.1 in \citeN{PlatenHe06} that the GOP achieves the largest average growth rate and, in the long run, pathwise the largest portfolio value. The benchmark is the central security of the benchmark approach; see \citeN{PlatenHe06} and \citeN{Platen26e}. \\
In \citeN{Platen26e}, stocks are modeled as portfolios of {\em factors}, which are independent auxiliary securities. The normalized factors and the normalized benchmark evolve in their respective activity times. However, market evolution is observed in calendar time, which complicates estimation and modeling. \citeN{Platen26e} uses the information-theoretic notion of {\em surprisal}, which is Shannon's entropy, to formulate assumptions that allow derivation of the core relationships of a financial market and lead to a realistic model for the benchmark. The following four assumptions or mathematical principles, formulated in \citeN{Platen26e}, underpin the proposed model for the benchmark, with the {\em activity} as the derivative of its {\em activity time}: \\

\noindent{\bf A1:} The  normalized factors are ergodic scalar diffusions driven by independent Brownian motions that evolve in  respective  activity times  with  quasi-stationary  derivatives.\\
	
	\noindent{\bf A2:} The market’s GOP and the unique solution to the system of stochastic differential equations characterizing the market dynamics exist. \\ 
	
		\noindent{\bf A3:} The surprisals of the quasi-stationary probability densities of the normalized factors and the normalized activity in their activity times  are maximized. \\
		
		\noindent{\bf A4:} The divergence between the BN pricing measure and the real-world probability measure is minimized.\\
		
	The benchmark approach does not rely on  the existence of expected-utility-maximizing representative agents or the absence of FLVRs. Its statements are mathematically derived from  the above four assumptions or principles and  stylized empirical facts.\\

As in \citeN{Platen26e}, we interpret the market  as a communication system; see \citeN{Shannon48} and \citeN{Kullback59}. In a surprisal-maximized market, distinct market dynamics are derived in \citeN{Platen26e}, where the factors, their sums, and the benchmark follow, in respective activity times, squared radial Ornstein-Uhlenbeck (SROU) processes, as studied, e.g., in \citeN{GoingYo03}. SROU processes extend the Cox-Ingersoll-Ross (CIR) process, which gained prominence through the work of \citeN{CoxInRo85}. Under the BN pricing measure, the benchmark follows a time-transformed squared Bessel process of dimension four in its activity time, which the current paper aims to model. \\
 In \citeN{Platen26e}, the squared volatility of the benchmark with respect to its activity time is predicted  to have a stationary density that is inverse gamma with four degrees of freedom. This  density   provides in activity time the highest possible rate of information transmission through the normalized benchmark as a communication channel. The current paper models the speed of this information transmission, which is the derivative of the activity time, the activity.\\ 
The  {\em MMM parametrization}, with the   activity  as state variable, follows from the surprisal maximization postulated in \citeN{Platen26e}. We employ it as  alternative to the Black-Scholes parametrization, which considers the  volatility as   state variable. The MMM parametrization provides, by construction,  an exact model for the benchmark in activity time, which leads to  neglidgible hedge errors; see, e.g., \citeN{Platen25a} and \citeN{PlatenFe25a}.\\ 
 The  activity time of the benchmark evolves approximately linearly and is stochastic, as shown in \citeN{PlatenRe26}. What is  missing is a theoretical understanding of the nature of the speed of information transmission, which the current paper identifies as a result of surprisal maximization and  overreaction in trading behavior. This leads to  a model for the fast-moving    activity of the benchmark that is useful for  BN pricing and hedging. The current paper constructs a proxy for the benchmark, derives  a respective ergodic scalar diffusion model for its normalized activity,  and compares predicted  properties with those empirically observed. According to \citeN{PlatenRe12an}, sufficiently diversified stock portfolios can be interpreted as proxies of the benchmark. 
 \\ 
  The following ten  stylized empirical facts extend those listed in \citeN{Platen26e} and will be shown to be consistent with the proposed model for  the benchmark:\\
 
 \noindent (1.)	Stock market index volatilities tend to follow processes with quasi-stationary probability densities, as demonstrated in \citeN{Engle82}.\\
 
 \noindent (2.)	Stock index log-return densities  approximate  Student t densities with slightly less than four degrees of freedom across various statistical methodologies in \citeN{MarkowitzUs96a}, \citeN{MarkowitzUs96b}, \citeN{HurstPl97d},  \citeN{FergussonPl06dc}, and \citeN{BreymannLuPl09e}.\\ \\
 (3.) Stock market indices show the leverage effect: declines in the index are associated with increased volatility, and vice versa; as observed, e.g., in \citeN{Black76b}, \citeN{AitSahalia12}, and \citeN{ElEuchFuRo18}.\\ \\
 (4.) Volatilities show rough  paths with occasional spikes that differ from typical trajectories of  diffusion processes; see, e.g., \citeN{BayerFrGa16}, and \citeN{ElEuchFuRo18}.\\ \\
 (5.) Volatilities comprise both rapidly and slowly changing components, as observed and modeled, e.g., in \citeN{FouquePaSi99}.\\ \\
 (6.) Log-returns of various stock portfolios  exhibit scaling properties and self-similarity for a wide range of observation frequencies, as evidenced by \citeN{Mandelbrot01a} and \citeN{BreymannLuPl09e}.\\ \\
 (7.) The logarithms of  stock market indices demonstrate approximately linear growth with mean-reverting characteristics; exemplified, e.g.,  in Figure \ref{FigK01}.\\ \\
 (8.) The covariation of the driving Brownian motions of a stock index and its volatility appears to be on average zero, as observed in \citeN{BreymannKePl06s}, which is called leverage effect puzzle in \citeN{AitSahalia12}. \\ \\
 (9.) For high-frequency log-returns of  a world stock index, the estimated density appears to be close to a variance gamma density with low degrees of freedom; see \citeN{BreymannKePl06s} and \citeN{BreymannLuPl09e}.\\ \\
 (10.) The diffusion coefficient with respect to calendar time of the activity of a stock  index appears  to be proportional to the activity; see \citeN{BreymannKePl06s}.\\ 
 
 To allow accurate benchmark-neutral pricing and hedging, a \textquoteleft realistic'  model for the benchmark, as referenced in this paper and based on the above four principles, should provide robust explanations for the above list of empirical facts.\\

The paper is organized as follows: Section 2 describes the benchmark dynamics in  activity time and introduces the MMM parametrization. The dynamics of the activity  are modeled and illustrated in Section 3. Section 4 constructs a proxy for the benchmark. In Section~5 the log-return densities of various stock portfolios are estimated and the listed empirical properties are compared to those predicted by the proposed model.

\section{Surprisal-Maximized Benchmark Dynamics}

\subsection{Minimal Market Model Parametrization}

Continuous-time finance,  advanced by Merton in \citeN{Merton90}, has used the Black-Scholes model (BSM) as the standard framework for modeling risky securities, as described in \citeN{BlackSc73}. This model assumes the asset prices to follow a geometric Brownian motion, defined as the exponential of a drifted, time-transformed Brownian motion. By design, the BSM parametrization focuses on the volatility process as central state variable. Research has shown that volatility is inherently stochastic rather than deterministic, where some of its  empirical properties are  listed in the introduction. \\

The maximization of the surprisal of the stationary density of the normalized benchmark in \citeN{Platen26e} indicates that the popular BSM parametrization, which focuses on  stochastic volatility modeling, is not ideally suited for  modeling  the dynamics of the benchmark. The results in \citeN{Platen26e} suggest the {\em MMM parametrization}, which draws inspiration from a model introduced in \citeN{Platen01a} and studied in \citeN{Platen02g} and \citeN{PlatenHe06}. This parametrization uses the {\em minimal market model} (MMM) from \citeN{Platen01a} to parametrize stock portfolio dynamics. It characterizes the theoretical surprisal-maximizing  dynamics of the benchmark in its {\em activity time}. \\

 Let us explain the MMM parametrization for the continuous benchmark $S^*_{t}$ at the calendar time $t\in[0,\infty)$ in savings account denomination.
 The activity time $\tau_t$ is given by the formula
 \begin{equation}\label{taut1}
	\tau_t=\ln\left([\sqrt{S^*_.}]_{t_0,t}+e^{\tau_{t_0}}\right)
\end{equation} 
for $t\in[t_0,\infty)$. The quadratic variation $[\sqrt{S^*.}]_{t_0,t}$ is approximated by summing the squares of the increments of $\sqrt{S^*_t}$ over a calendar-time discretization with sufficiently  small step size. More precisely, the quadratic variation $[\sqrt{S^*_.}]_{t_0,t}$ equals the $P$-almost sure limit of the sum of the squares of the increments of $\sqrt{S^*_t}$; see, e.g., \citeN{PlatenHe06}.\\ Without loss of generality, let us assume that the square root $\sqrt{S^*_t}$ of a continuous portfolio $S^*_t$ satisfies the stochastic differential equation (SDE)
 \begin{equation}
	d\sqrt{S^*_t}=\mu_{t}dt+\sigma_{t}d\tilde W_{t}.
\end{equation}
In this context, $\tilde W_{t}$ denotes the driving Brownian motion of $S^*_t$ with respect to the calendar time  $t \in [t_0, \infty)$, while $\mu_{t}$ and $\sigma_{t}$ represent general continuous processes that allow the subsequent manipulations in a given filtered probability space $(\Omega,\mathcal{F},\underline{\cal{F}},P)$ with filtration $\underline{\cal{F}}$ $=(\mathcal{F}_t)_{t \in [{t_0},\infty)}$ satisfying the usual conditions, as described, e.g., in \citeN{PlatenHe06}.
\\

 Motivated by the MMM, which constitutes the surprisal-maximizing dynamics of the benchmark in activity time, as derived in \citeN{Platen26e}, the MMM parametrization uniquely specifies the portfolio's activity time ${\tau_t}$ via the differential equation
\begin{equation}\label{dtaut}
	\frac{d{{\tau_t}}}{dt}=\sigma^2_te^{-{\tau_t}}
\end{equation}
for $t\in [t_0,\infty)$ with  initial  activity time ${\tau_{t_0}}$. By application of the It\^o formula it follows for $S^*_t$ the SDE
\begin{equation}
	dS^*_t
	=\left(2\sqrt{S^*_t}\mu_{t}+\sigma^2_t\right)dt+\sqrt{S^*_t4\sigma^2_t}d\tilde W_{t}.
\end{equation} 
One notices by \eqref{dtaut} that it holds \begin{equation}
	\sigma^2_t=e^{{\tau_t}}\frac{d{\tau_t}}{dt},
\end{equation}
 which  allows one to rewrite the above SDE in the form
\begin{equation}\label{dS*}
	dS^*_t
	=\left(2\sqrt{S^*_t}\mu_{t}+\sigma^2_t\right)dt+\sqrt{S^*_t4e^{{\tau_t}}}dW_{\tau_t},
\end{equation} 
where $W_{\tau_t}$ denotes the respective Brownian motion in activity time $\tau_t$ with stochastic differential
\begin{equation}
	dW_{\tau_t}=\sqrt{\frac{d\tau_t}{dt}}d\tilde W_t
\end{equation}
and $W_{\tau_{t_0}}=0$.
The diffusion coefficient of this SDE takes the  form of that of the MMM, which follows from assuming \eqref{dtaut}. The MMM parametrization involves the activity time ${\tau_t}$, the benchmark $S^*_t$, and the driving Brownian motion $W_{\tau_t}$ in activity time. The diffusion coefficient of almost every strictly positive continuous process can be parametrized using the MMM parametrization, yielding an observable activity time.  In the current paper, the benchmark (or proxy of the benchmark) is modeled as continuous process. In reality, the process $S^*_t$ may have jumps. These jumps generate upward jumps in the activity time. Despite this, the MMM parametrization still makes good sense, as we will see.\\

\begin{figure}
 	\includegraphics[width=\textwidth]{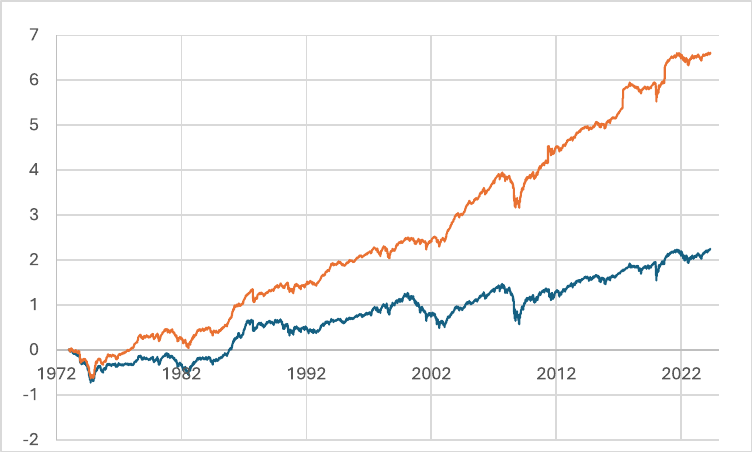}
 	\caption{Logarithms of MSCI (in blue) and EWI163 (in red) in US savings account denomination.}
 	\label{FigK01}
 \end{figure}

\begin{figure}
 	\includegraphics[width=\textwidth]{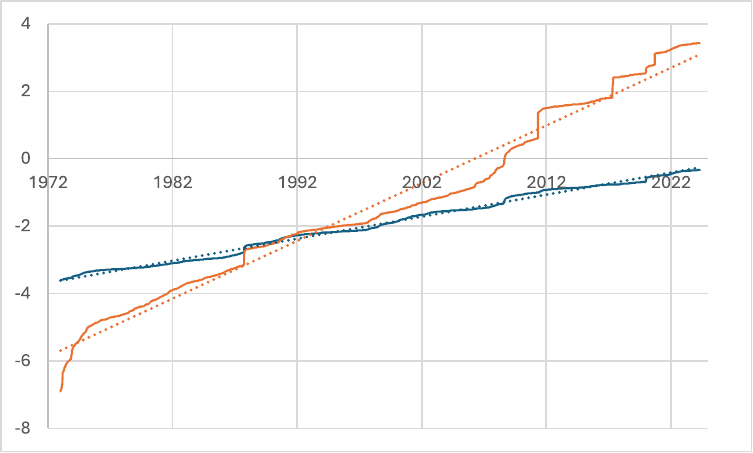}
 	\caption{Activity times of MSCI (in blue) and EWI163 (in red) with trendlines.}
 		\label{FigK03}
 	\end{figure}

For illustration, let us apply the MMM parametrization to a well-known benchmark employed in fund management, the daily observed MSCI total return world stock index (MSCI)  in US savings account denomination. 
 Figure \ref{FigK01} displays the logarithm of the MSCI in blue, which has an average annualized growth rate of $0.05$.    This figure displays also the logarithm of the EWI163 in red, which we construct in Section \ref{sec:methodology} as proxy for the benchmark. It has an average annualized growth rate of $0.139$, which is almost $8.9$\% higher than that of the MSCI.\\ The estimated activity times of the MSCI (in blue) and the EWI163 (in red) are shown in Figure \ref{FigK03} together with their trendlines.
The initial activity time is estimated as follows: First, a tentative initial activity time is chosen and Equation \eqref{taut1} applied to calculate the  activity time. For proxies of the benchmark, the activity time typically exhibits an approximately linear pattern, indicating that it evolves in an approximately linear manner, as can be seen in Figure \ref{FigK03}. To capture this pattern, standard linear regression is applied to fit a trendline. The estimate of the initial activity time $\tau_{t_0}$ is obtained as the starting value of the activity time that produces the highest $R^2$ score. For the activity time of the MSCI, displayed in Figure~\ref{FigK03} in blue, the initial activity time is estimated as $-3.6$, the slope of its trendline amounts to $0.065$, and the $R^2$ score   to $0.993$, indicating that the activity time follows approximately a linear pattern. The activity time of the EWI163, displayed in red in Figure \ref{FigK03}, has the initial activity time of $-6.9$, the slope $0.171$ of its trendline, and  an $R^2$ score of $0.976$. 
One notices a small jump in the activity of the MSCI during the 1987 Stock Market Crash, when some markets were closed for several days, and a visible more-than-usual increase during the Global Financial Crisis in 2007/2008. The activity time of the EWI163 is increasing faster but pays with a few jumps and a lower $R^2$ score for the higher slope of its trendline. As explained in \citeN{Platen26e}, jumps of the benchmark  occur theoretically when factors approach zero, which makes the modeling and quantitative analysis challenging. The model and quantitative methods proposed in the current paper are designed to deal  with occasionally occurring  jumps. \\
 \begin{figure}
	\includegraphics[width=\textwidth]{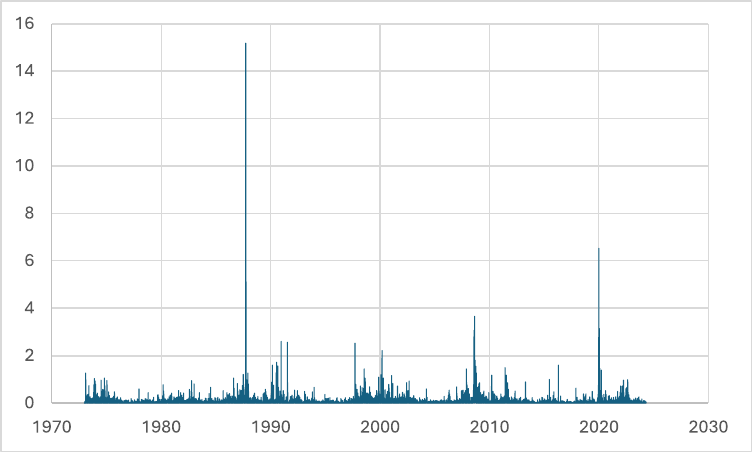}
	\caption{Activity of MSCI in calendar time.}
	\label{Figure13}
\end{figure}
We display in Figure \ref{Figure13} the activity of the MSCI in calendar time, which we obtain as numerical derivative of its activity time with respect to the average activity time. 
\begin{figure}
	\includegraphics[width=\textwidth]{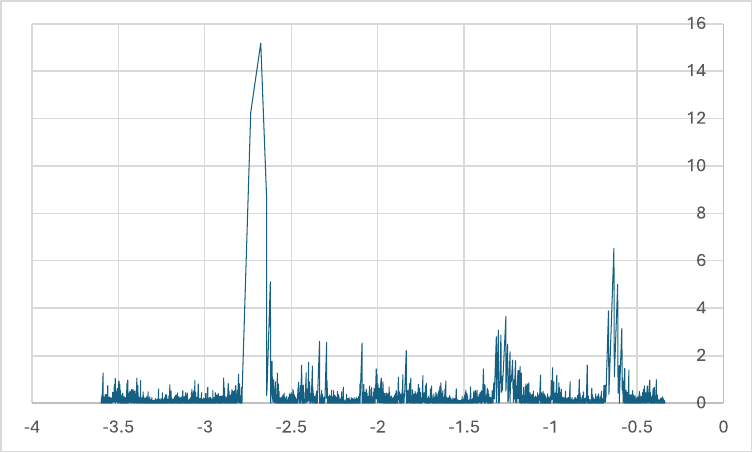}
	\caption{Activity of MSCI in activity time.}
	\label{Figure14}
\end{figure}For comparison, we display in Figure \ref{Figure14} this activity in activity time. One notices that in the latter figure the activity appears   likely to follow  a fast moving diffusion process besides during the 1987 Stock Market Crash. One notes that spikes and roughness of the activity in Figure \ref{Figure13} are mostly due to its observation in calendar time. When various jumps are present in the trajectory of the benchmark, as for the EWI163, the jumps visually dominate  the graph and the respective activity time appears to be more informative.
\subsection{Benchmark-Neutral Pricing}

 The benchmark-neutral (BN) pricing measure $Q_{S^*}$, introduced in \citeN{Platen25a}, is the pricing measure that arises when the stock portfolio $S^*_t$ is used as the numéraire, under the assumption that $S^*_t$ serves as  proxy for the benchmark. In \citeN{Platen25a}, \citeN{Platen26a}, and \citeN{SchmutzPlSc25}, BN pricing is shown to provide a valuation method that can realistically replace risk-neutral pricing. For short terms to maturity, BN pricing yields prices similar to formally obtained risk-neutral prices. For longer terms to maturity, BN prices of nonnegative contingent claims can be significantly lower than risk-neutral ones, as demonstrated, e.g., in \citeN{Platen25a} and \citeN{PlatenFe25a}.\\

For BN pricing, the drift of the SDE \eqref{dS*} does not play a crucial role, as follows from \citeN{Platen25a}. The above-described MMM parametrization of the diffusion coefficient in the SDE \eqref{dS*} is well suited for BN pricing because it yields a tractable and realistic model for the benchmark in activity time under the BN pricing measure. The activity time is observable and the drift and diffusion coefficients of the model dynamics are  known. To facilitate BN pricing, one models the benchmark dynamics in savings account denomination by introducing the drifted Brownian motion $\bar W_{\tau_t}$, which is given by the stochastic differential
\begin{equation}
	d\bar W_{\tau_t}=\frac{(2\sqrt{S^*_{t}}\mu_{t}+\sigma^2_t)dt-4e^{{\tau_t}}d{\tau_t} }{\sqrt{S^*_{t}4e^{{\tau_t}}}}+dW_{\tau_t}.
\end{equation}
When $\bar W_{\tau_t}$ is assumed to be a Brownian motion evolving in the  activity time ${\tau_t}$ under the BN pricing measure $Q_{S^*}$, the SDE
\begin{equation}\label{S*}
	dS^*_t=4e^{{\tau_t}}d{\tau_t}+\sqrt{S^*_t4e^{{\tau_t}}}d\bar W_{\tau_t}
\end{equation} 
emerges as that of a time-transformed squared Bessel process with dimension four, which constitutes the MMM in the activity time ${\tau_t}$; see \citeN{RevuzYo99}, \citeN{Platen01a}, and \citeN{Platen25a}. This SDE for the savings account-denominated benchmark $S^*_t$ in activity time $\tau_t$  under the BN pricing measure $Q_{S^*}$ is predicted by the surprisal-maximized market dynamics derived in \citeN{Platen26e}. In reality, the activity time evolves approximately linearly in calendar time, as shown in Figure \ref{FigK03}.\\ For a nonnegative contingent claim $H_T$ with bounded stopping time $T$ and $\mathcal{F}_T$-measurable payoff satisfying the integrability condition
\begin{equation}
	{\bf{E}}^{Q_{S^*}}\left(\frac{H_T}{S^*_T}\bigg|\mathcal{F}_t\right)<\infty,
\end{equation} the {\em BN pricing formula}
provides according to \citeN{Platen25a} at the time $t\in[t_0,T]$ the price
\begin{equation}\label{BNPrice}
H_t=	S^*_t{\bf{E}}^{Q_{S^*}}\left(\frac{H_T}{S^*_T}\bigg|\mathcal{F}_t\right).
\end{equation}
For a contingent claim whose payoff depends only on the value of the benchmark at a fixed maturity in activity time, the BN probability density of the payoff is given by the non-central chi-squared transition density of the benchmark under the MMM. Since the activity time can be observed as it evolves, the  MMM dynamics are perfectly known in activity time for the benchmark under the BN pricing measure even in the case when the proxy of the benchmark and its activity time have jumps. Therefore, using the BN pricing formula \eqref{BNPrice}, one can determine exactly the BN price of the contingent claim. The respective hedge can be expected to work almost perfectly when the maximum step size of the activity time discretization is sufficiently small and the jumps are not too high, as demonstrated in \citeN{Platen25a} and \citeN{PlatenFe25a} for extreme-maturity zero-coupon bonds that pay one unit of the savings account for maturities that reach over several decades. Thus, for  sufficiently small  activity time steps and reasonable jump sizes in the  benchmark, the resulting hedge error can be expected to be negligible. When  jumps occur in the benchmark $S^*_t$ or the step size in activity time  becomes occasionally  large, then one has to expect slightly larger hedge errors, as have been observed in \citeN{Platen25a} and \citeN{PlatenFe25a}. Important is the fact that for proxies of the benchmark with higher average growth rate, the BN prices of nonnegative contingent claims, like those of   zero-coupon bonds, are, on average, smaller. This means that it is advantageous to perform BN pricing with a proxy of the benchmark that has a high average growth rate, like the EWI163.\\

If the maturity of a contingent claim is not a fixed activity time but rather a given calendar time, then one needs to model the   activity time so that the valuation in the BN pricing formula \eqref{BNPrice} accounts for the randomness of the activity time at maturity.

\subsection{Stationary Density of the Normalized Benchmark}

The SDE \eqref{dS*} for the benchmark shows that the MMM dynamics in activity time capture the leverage effect, listed as empirical fact (3.), which arises as follows: When the stock portfolio decreases, its squared volatility  \begin{equation}\label{sqvol}
	\frac{4e^{\tau_t}}{S^*_t}=\frac{2}{Y_{\tau_t}}\end{equation}
	 with respect to activity time increases, and vice versa. Under the MMM and the BN pricing measure, the It\^o formula implies that  the normalized benchmark
 \begin{equation}
	Y_{\tau_t}=\frac{S^*_t}{2e^{\tau_t}},
	\end{equation}
	satisfies under the BN pricing measure the SDE
	\begin{equation}
		dY_{\tau_t}=(2-Y_{\tau_t})d\tau_t+\sqrt{2Y_{\tau_t}}d\bar W_{\tau_t}
	\end{equation}
	for $t\in[t_0,\infty)$. Therefore, the squared volatility with respect to activity time equals two times the inverse $\frac{1}{Y_{\tau_t}}$ of a Cox-Ingersoll-Ross (CIR) process $Y_{\tau_t}$ with mean $2$ and dimension $4$.	
 When considered with respect to activity time under the BN pricing measure, the stationary density of the CIR process $Y_\tau$ is a gamma density with four degrees of freedom and mean $2$. The squared volatility of the normalized benchmark in activity time equals, according to \eqref{sqvol}, the ratio $\frac{2}{Y_{\tau}}$, which has, in activity time $\tau$, a stationary density that is an inverse gamma density with four degrees of freedom. The dimension four of the CIR process $Y_\tau$ is the same under the real-world probability measure $P$ and the BN pricing measure $Q_{S^*}$; see \citeN{Platen25a}.  It follows from the information-theoretic  assumptions formulated in the introduction that the gamma density with four degrees of freedom represents  the stationary density with the maximum possible surprisal of the stationary density of the information channel provided by the  normalized benchmark.\\ When estimating the probability density generated by the log-returns of the normalized benchmark with respect to activity time, the estimated log-return density can be expected to follow a normal variance-mixture density, where the mixing density is an inverse gamma density with four degrees of freedom. 
 Since under the real-world probability measure $P$ only the mean of the above BN log-return density changes, this implies   that the estimated log-return density with respect to activity time is, for  the benchmark, predicted to be a Student $t$ density
  with four degrees of freedom. It follows  that the Student $t$ density with four degrees of freedom represents  the log-return density generated by the maximum possible surprisal of the stationary density of the information channel provided by the  normalized benchmark. For  large calendar time step sizes, like monthly and daily observations, most of the stochasticity of the activity is averaged out and one should expect, due to the randomness of the activity and potential jumps of the benchmark, slightly more leptokurtic log-returns, which suggests an estimated Student $t$ density with  slightly lower degrees of freedom between three and four. This  predicts, as estimated probability density of the benchmark's log-returns, a Student $t$ density with  slightly less than four degrees of freedom, which is consistent with the listed empirical facts (1.) and (2.). \\
   As predicted in \citeN{Platen26e}, the benchmark and various other stock portfolios show, in activity time, scaling and self-similarity properties as evidenced by the listed empirical fact (6.). Since the activity is modeled as a quasi-stationary process, the logarithms of proxies of the benchmark are predicted to evolve in the long-term approximately linearly, which has been listed as empirical stylized fact (7.).

 \section{Activity Model}	\label{marketactivity1}

 \subsection{Overreactions in Trading Behavior}

 Due to the SDE \eqref{S*}, the fast-moving   activity $a_t=\frac{d\tau_t}{dt}$ appears in the squared volatility 
 \begin{equation}\label{squvola}
 	\frac{4e^{\tau_t}\frac{d\tau_t}{dt}}{S^*_t}=\frac{2a_t}{Y_{\tau_t}},
 \end{equation} of the benchmark with respect to the calendar time $t\in[t_0,\infty)$.  
  The time-dependent average of the activity $a_t$ we model by the externally given {\em average activity} $\bar a_t$. The {\em normalized activity time} $ \bar\tau_t$ is given by the integral
  \begin{equation}
  	\bar \tau_t=\bar\tau_{t_0}+\int_{t_0}^{t}a(\bar\tau_s)ds,
  \end{equation}
  where 
  \begin{equation}\label{activityt}
  a(\bar\tau_t)=\frac{	a_t}{\bar a_t} 
  \end{equation} denotes the  {\em normalized activity} for $t\in[t_0,\infty)$.
   The average activity can capture seasonal patterns in activity, which typically depend on the time of day, day of the week, week of the year, and macroeconomic cycles. For simplicity, we will set the average activity in the remainder of the paper equal to the constant $\bar a=\bar a_t>0$ because this simplification appears to be sufficient when considering daily data. The normalized activity is modeled as an ergodic scalar diffusion in normalized activity time. It follows from the information-theoretic  assumptions formulated in the introduction that a gamma density  represents  the stationary density with the maximum possible surprisal of the stationary density of the information channel provided by the  normalized activity.\\

 To identify a respective model for the activity that is consistent with the listed empirical facts and Assumption A3, one has to model realistically   the timing of  trades. Traders use moving averages and bands around them as inputs for trading algorithms. Examples include the Keltner channels described in \citeN{Keltner60} and Bollinger bands, explained, e.g., in \citeN{Baiynd11}. The purpose of Keltner channels and Bollinger bands is to provide a relative definition of high and low prices. Prices are considered high at the upper band and low at the lower band, where the distance between these bands typically relates to the average  deviation of returns. These bands are widely used to inform systematic trading decisions, which determine the speed of price information transmission.  There exist different groups of market participants that   use different observation frequencies to trigger trades.  We model, for simplicity, the trading behavior of one representative group and restrict our modeling to a respective step size  where the diffusion approximation of the normalized activity  makes sense. One can refine the proposed model by differentiating between trades of groups of market participants.\\ 
 
  Let us introduce a discretization $\bar\tau_{t_0},\bar\tau_{t_1},\bar\tau_{t_2},...$ of the  normalized activity time. We denote by $W_{\bar\tau_{t_i}}$ the value of the Brownian motion $W_{\bar\tau_{t}}$ with respect to the normalized activity time $\bar\tau_t$ that is driving the proxy of the benchmark at the $i$-th discretization time. 
   Similarly as for Keltner channels and Bollinger bands,  we introduce
  a {\em moving average} $U(\bar\tau_{t})$ of $W_{\bar\tau_t}$. The following algorithm yields at the $i$-th discretization time $t_i$, $i\in\{1,2,...\}$, the value of the moving average $U(\bar\tau_{t_i})$ in the form
   \begin{equation} \label{AAAdYtilde0} U(\bar\tau_{t_i})=\kappa W_{\bar\tau_{t_i}}
  	+(1-\kappa)U(\bar\tau_{t_{i-1}})\end{equation}
     for $i\in\{1,2,...\}$ with $U(\bar\tau_{t_0})=W_{\bar\tau_{t_0}}$,  where $\kappa$ is  the {\em smoothing parameter}.
   One  facilitates for $\kappa\in (0,1)$  some smoothing of the trajectory of the driving Brownian motion $W_{\tau_{t_i}}$. For $\kappa$ close to $1$, one  approximates closely  the Brownian motion path.  
   
      By adding and subtracting, one can construct  from \eqref{AAAdYtilde0} the equation
   \begin{equation} \label{AAdYtilde01} \frac{U(\bar\tau_{t_i})+U(\bar\tau_{t_{i-1}})}{2}=\frac{\kappa}{2}W_{\bar\tau_{t_i}}
   	+\left(1-\frac{\kappa}{2}\right)U(\bar\tau_{t_{i-1}})\end{equation}
   for $i\in\{1,2,...\}$.
    In numerical analysis, this algorithm resembles a predictor-corrector or implicit numerical scheme, which one typically uses to control the propagation of discretisation errors by forcing the approximation to remain near the solution; see, e.g., Chapter 12 in \citeN{KloedenPl92}.  When $\kappa$ is between $1$ and $2$, the algorithm \eqref{AAAdYtilde0}  models  {\em overreactions} in the trading behavior.  In this case,   when the Brownian motion moves more than usual, the trajectory of the moving average does not smooth these moves but instead exaggerates them. The moving average  fluctuates still around the Brownian motion path in this case. For a smoothing parameter $\kappa \geq 2$ the  moving average would eventually diverge from the Brownian motion path.\\ From the perspective of trading behavior,  the case $\kappa\in(1,2)$ appears to be  realistic because an extremely large number of market participants is reacting in almost the same way to the evolving Brownian motion, which  generates overreactions when the Brownian motion moves more than usual. We assume in this paper that the market participants act continuously with   overreactions in trading behavior  to larger-than-usual moves of the Brownian motion. The above algorithm facilitates the modeling of these overreactions by assuming $\kappa\in(1,2)$. These overreactions manifest as random fluctuations of the moving average around the Brownian motion path.    The overreactions in trading behavior are becoming larger and longer-lasting when the smoothing parameter $\kappa$ gets closer to $2$ but remains below $2$.  \\
    \begin{figure}
   	\includegraphics[width=\textwidth]{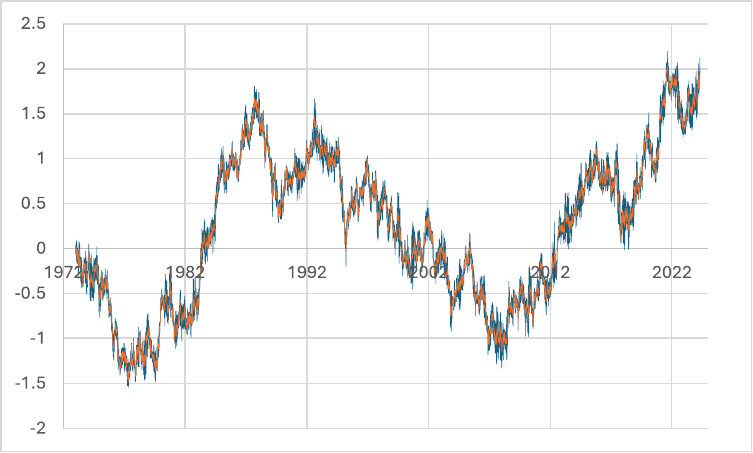}
   	\caption{Brownian motion(in red) and moving average (in blue).}
   	\label{Figure15}
   \end{figure}
   
  The normalized activity   is observed in \citeN{BreymannKePl06s} to be higher  during periods when the Brownian motion path moves more than usual. It, therefore, reaches more frequently  the boundaries of  Keltner channels and Bollinger bands. To capture the resulting trading behavior quantitatively, similarly as the squares of the distances between the bands of Keltner channels and Bollinger bands are chosen, we specify the  normalized activity  as the functional
   \begin{equation} \label{AdMt0} a(\bar\tau_{t_i}) 
   	=\xi \left(\frac{U(\bar\tau_{t_i})-U(\bar\tau_{t_{i-1}})}{\bar\tau_{t_{i}}-\bar\tau_{t_{i-1}}}\right)^2,
   \end{equation}
   where   $\xi>0$ is called  {\em scaling parameter}. 
     For illustration, we show in Figure \ref{Figure15} a simulated Brownian motion path (in red)  for $\kappa=1.97$ and $\bar a=0.15$ together with the moving average (in blue) for the  calendar time step size $\Delta=1/262$. We use here the estimated parameters of the daily oberved EWI163 and its time axis for performing the illustrating simulation. One notes the captured overreaction effect in Figure \ref{Figure15}, where the moving average fluctuates around the Brownian motion path. When the smoothing parameter is very close to $2$, the normalized activity in calendar time appears to consist almost only of large spikes. Spikes are also generated by jumps in the benchmark. Therefore, by an appropriate choice of $\kappa$ close to $2$ one can simulate  the effect of occasional  jumps of the benchmark on maturities that represent fixed calendar times.
   \subsection{Normalized Activity as Ergodic Scalar  Diffusion}
 
   Even though trading is performed at discrete times,   the study of properties of stochastic processes appears to be easier when using continuous-time stochastic calculus with its power and elegance. It simplifies the derivations, keeps track of important effects that the It\^o formula captures, and allows us to identify stationary densities of stationary scalar diffusions. However, modeling  the market microstructure is complex and difficult, as described in \citeN{Ohara95}. Therefore, we search for an  ergodic scalar diffusion as an approximate model for the normalized activity, which requires only  identification of its drift and diffusion coefficients with respect to the intrinsic activity time. We know already from Assumption A3 that the stationary density of a diffusion representing an information channel must be a gamma density because this is the surprisal-maximizing stationary density of a diffusion under analogous constraints as the normalized factors, which represent  information channels.\\
    For a sufficiently small calendar-time  step size $\Delta>0$, we can write the algorithm  \eqref{AAAdYtilde0} in the form
    \begin{equation} \label{1AdYtilde0} U(\bar\tau_t)-U(\bar\tau_{t-\Delta})=\kappa_\Delta\left(W_{\bar\tau_t}
  	-U(\bar\tau_{t-\Delta})\right)(\bar\tau_t-\bar\tau_{t-\Delta})\end{equation}
  for $t\in[t_0+\Delta,\infty)$ with $\kappa_\Delta=\frac{\kappa}{\bar a \Delta}$. By following  well-established arguments, which secure the discrete time approximation of solutions of SDEs as described in Chapter 9 of \citeN{KloedenPl92}, we characterize approximately the moving average by the SDE
        \begin{equation} \label{AdYtilde0} dU(\bar\tau_t)\approx\kappa_\Delta\left(W_{\bar\tau_t}
 	-U(\bar\tau_t)\right)d\bar\tau_t\end{equation}
 for $t\in[t_0,\infty)$ with $U(\bar\tau_{t_0})=W_{\bar\tau_{t_0}}$, where $\kappa_\Delta\in (0,\frac{2}{\bar a \Delta})$ is called  the {\em speed of adjustment}. This leads us by \eqref{AdMt0} to the  normalized activity in the form \begin{equation} \label{1AdMt0} a(\bar\tau_{t}) 
 	\approx\xi \left(\frac{dU(\bar\tau_t)}{d\bar\tau_{t}}\right)^2.
 \end{equation} 
 This activity model has similarities with the activity  model proposed in \citeN{PlatenRe26}. However, the latter  smoothes   a different process in calendar time, involves a base activity, and does not identify an approximating ergodic scalar diffusion.
 	 By application of the It\^o formula, the above normalized  activity  satisfies the SDE
  \begin{equation}
 	d a(\bar\tau_t)\approx\xi\kappa_\Delta^2\left(2(W_{\bar\tau_t}
 	-U(\bar\tau_t))(dW_{\bar\tau_t}			-\kappa_\Delta(W_{\bar\tau_t}
 	-U(\bar\tau_t))d\bar\tau_t)+d\bar\tau_t\right).
 \end{equation}
 This yields by \eqref{1AdMt0} the SDE
 \begin{equation}\label{1daDeltat}
 	d a(\bar\tau_t)\approx 2\kappa_\Delta\left(\frac{\xi \kappa_\Delta}{2} - a(\bar\tau_t)\right)d\bar\tau_t+\sqrt{\xi}2\kappa_\Delta\sqrt{a(\bar\tau_t)}\operatorname{sgn}(\bar\tau_t)dW_{\bar\tau_t}
 	,
 \end{equation}
 with sign function
 \begin{equation}\label{AAequ2115''}
 	\operatorname{sgn}(\bar\tau_t)=\frac{	W_{\bar\tau_t}
 		-U(\bar\tau_t)}{|W_{\bar\tau_t}
 		-U(\bar\tau_t)|}
 \end{equation}
 for $W_{\bar\tau_t}
 -U(\bar\tau_t)\neq0$ and $\operatorname{sgn}_{{\bar\tau_t}}=0$ for $	W_{\bar\tau_t}
 -U(\bar\tau_t)=0$   for $t \in [{t_0}, \infty)$. 
  In normalized  activity time, this is  the SDE of the CIR process $a(\bar\tau_t)$, which has dimension $1$ and mean $\frac{\xi \kappa_\Delta}{2}$.
  Since this process is much faster moving than the normalized benchmark, it is consistent with the listed empirical fact (5.). \\ For high-frequency log-returns, the derived normalized activity  model can be expected to determine the normal variance-mixture density of the log-returns of the proxy of the benchmark when observed in normalized activity time. As will be explained in Section \ref{logreturn}, the stationary gamma density of the normalized activity with one degree of freedom  determines in this case the log-return density of the benchmark as a variance-gamma density with one degree of freedom. This theoretically predicted property is consistent with the listed empirical fact (9.).\\
  
   Furthermore, when maximizing the surprisal of the stationary density of an ergodic scalar  diffusion under the constraints that its arithmetic and logarithmic averages represent given parameters,  a gamma density results as the optimal density, as shown in \citeN{Platen26e} for the case of the stationary density of the normalized benchmark. As postulated in A3, the stationary density of the normalized activity is assumed to yield maximum surprisal. The degree of freedom $1$ of this gamma density follows from the fact that one models here the impact of a single Brownian motion on the activity dynamics.\\ For less frequent observations, some random effects of the fast-evolving normalized activity become averaged and the stationary density of the inverse of the normalised benchmark determines primarily the normal variance-mixture density of the log-returns, which are by   A1, A2, and A3 in \citeN{Platen26e} shown to be Student $t$ distributed with slightly less than four degrees of freedom, as  will be discussed and tested  in Section \ref{sec:SGH}.\\
  \begin{figure}
  	\includegraphics[width=\textwidth]{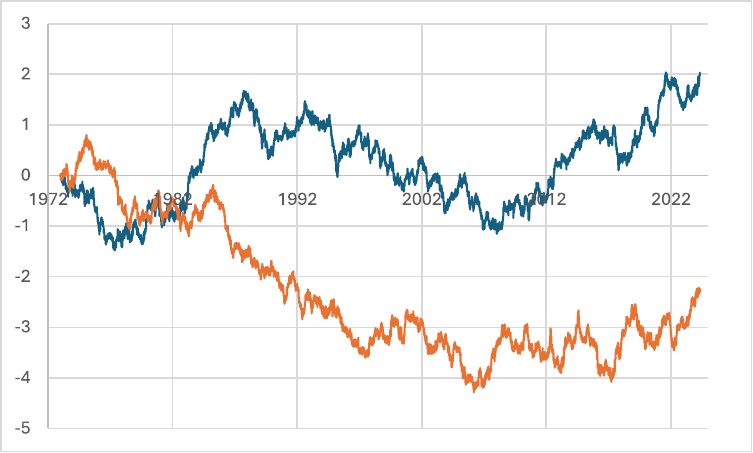}
  	\caption{Brownian motions driving the benchmark (blue) and the normalized activity (red).}
  	\label{Figure16}
  \end{figure}
    \begin{figure}
    	\includegraphics[width=\textwidth]{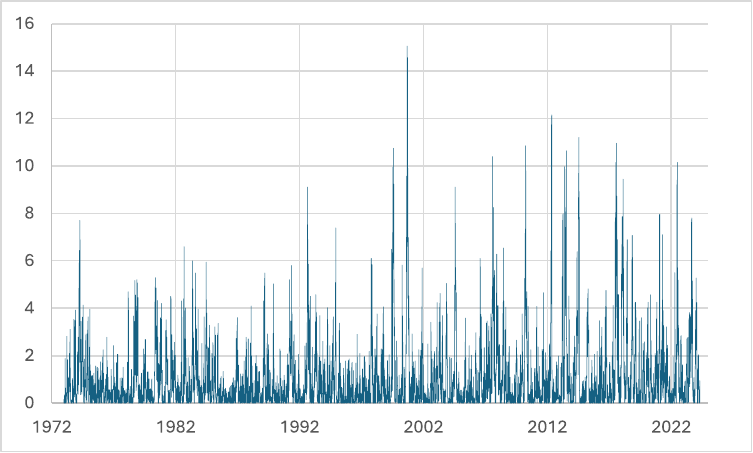}
    	\caption{Simulated normalized activity in calendar time.}
    	\label{Figure17}
    \end{figure}
    \begin{figure}
    	\includegraphics[width=\textwidth]{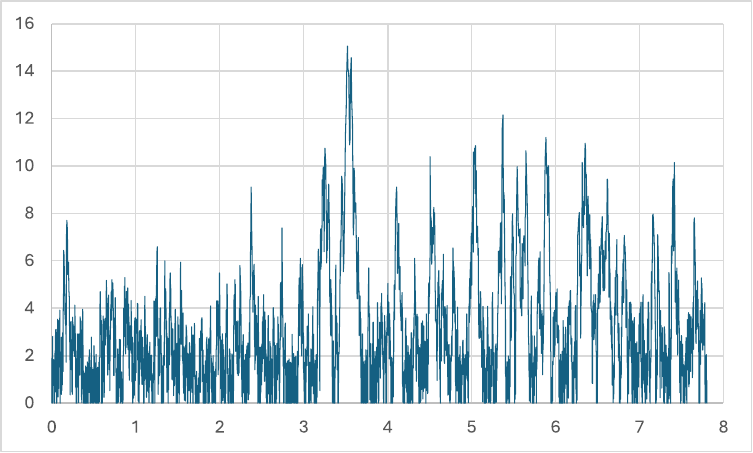}
    	\caption{Simulated normalized activity in normalized activity time.}
    	\label{Figure18}
    \end{figure}
  
The sign function, given in \eqref{AAequ2115''}, transforms  the Brownian motion driving the SDE of the proxy of the benchmark into the Brownian motion driving its  normalized activity. 
Since the sign function switches frequently in a symmetric manner, the covariation between these two Brownian motions is, under the proposed model, predicted to vanish on average. This  explains the listed empirical fact (8.). For illustration, Figure \ref{Figure16} shows the simulated Brownian motion $W_{\bar\tau_t}$ in blue and the Brownian motion that drives the normalized activity in red by including the sign function.  If one considers the diffusion coefficient of the normalized activity in calendar time, then it is proportional to the normalized activity, which  makes it consistent with the empirical fact (10.).\\
We plot in Figure \ref{Figure17} the simulated normalized activity in calendar time and in Figure \ref{Figure18} its trajectory in normalized activity time from our  simulation for $\xi=0.0042$, $\kappa=1.97$ and $\bar a=0.15$.  One notices that the normalized activity has in calendar time a rougher trajectory and spikes, which appear in normalized activity time less pronounced and the spikes seem to be caused by excursions of the moving average. These predictions are consistent with the listed empirical fact (4.). When bringing the value of the smoothing parameter $\kappa$ even closer to $2$,  one obtains a normalized activity in calendar time that consists visually almost entirely of spikes. This means that one can simulate the effect of jumps in the benchmark on the normalized activity time by setting $\kappa$ sufficiently close to $2$ when pricing and hedging contingent claims using Monte Carlo simulation.\\
It is worth noting that the normalized activity can be expressed as the square $a(\bar\tau_t)=(b(\bar\tau_t))^2$
 of the Ornstein-Uhlenbeck process $b(\bar\tau_t)$ that satisfies the SDE
 \begin{equation}
 	d b(\bar\tau_t)=-\kappa_\Delta b(\bar\tau_t)d\bar\tau_t+\sqrt{\xi}\kappa_\Delta dW_{\bar\tau_t}
 \end{equation}
for $t\in[t_0,\infty)$.
One can easily simulate trajectories of the normalized activity and the normalized benchmark in activity time using the above described algorithm together with numerical methods for SDEs  provided, e.g., in  Chapter 9 of \citeN{KloedenPl92}. Both processes originate from one Brownian motion. With the activity in activity time one obtains the respective calendar time and can identify maturities that are fixed in calendar time. Only a few parameters are needed  for BN pricing and hedging of contingent claims, which can be performed using Monte Carlo simulation with variance reduction, as proposed, e.g., in \citeN{HeathPl02a}. Extremely accurate  BN pricing  and hedging has been demonstrated  in \citeN{Platen25a} for extreme-maturity zero-coupon bonds with fixed activity times as maturities.

 \section{Proxies of the Benchmark}\label{sec:methodology}
 
 \subsection{Construction of Equal-Weighted Indices}
Since it is paramount for BN pricing and hedging to have access to  a good proxy of the benchmark, this section describes the construction of selected stock portfolios as equal-weighted indices (EWIs) to approximate the benchmark of the stocks. In \citeN{Platen26e}, stocks are assumed to be driven by a set of independent Brownian motions.  Each independent Brownian motion is assumed to drive a {\em  factor}, which   is a portfolio of stocks that is only driven by one independent Brownian motion. These factors form a stochastic basis and are auxiliary securities and not directly traded in the stock market. Theoretically, it has been shown in Theorem 3.6 in \citeN{Platen26e} that the equal-weighted factor portfolio represents the benchmark. \\To construct the factors, it would be necessary to estimate and invert an extremely large covariance matrix—a task that is  difficult to accomplish reliably in practice. Instead, in the current paper we approximate  factors by equally weighted groups of stocks that are, for economic reasons, driven  by similar sources of randomness. Equal-weighting  reduces the impact of idiosyncratic driving Brownian motions and highlights the common main driving Brownian motion of a  group of stocks.   \\
  The construction of EWIs in this paper is  similar to the methodology described in \citeN{PlatenRe12an}, where a proxy for the benchmark of the global stock market was constructed using global industry subsector indices. 
  We use the Global Industry Classification Standard (GICS), which  represents  a classification methodology that is applied to equity securities across the world's stock markets, developed  in 1999 by MSCI in collaboration with S\&P Dow Jones Indices. The  GICS is a registered trademark of McGraw Hill Financial and MSCI Inc. \\
 We employ four levels of classification in the GICS with respect to industries within a country, ranging from sectors, industry groups, industries, to sub-industries in increasing order of resolution. For example, as at August 2024, there are 11 sectors, these being: Energy, Materials, Industrials, Consumer Discretionary, Consumer Staples, Health Care, Financials, Information Technology, Communication Services, Utilities, Real Estate. Furthermore, there are 25 industry groups, 74 industries and 163 sub-industries, comprising a hierarchical classification; see \citeN{MSCI2024}.\\

 Our data consists of historical index values of all four levels of classification for 67 countries, commencing in January 1973 and finishing in July 2024, as provided by Thomson-Reuters. The index values represent the sum of the values of the stocks in a country in the respective sectors, industry groups, industries, and sub-industries. We
 denominate all index values in US Dollars and form the global  indices for each sector, industry group, industry, and sub-industry by equally weighting the corresponding indices of the   countries. We deliberately do not remove any  seemingly unusual  index moves or  jumps from the data provided
  for each day that the markets were open.\\ Denote by $t_i=t_0+i \Delta$, $i\in\{0,1,2,..\}$, the calendar time of the $i$-th observation, where $\Delta>0$ denotes the average calendar time step size. Let $S^{j,k}({t_i})$ denote the total return value of the market capitalisation-weighted index of the stocks of the $j$-th sector, industry group, industry, or sub-industry for the $k$-th country at time $t_i$, as obtained from Thomson-Reuters. 
 This means, we construct global EWIs at each level by first equally weighting across countries and then equally weighting over the resulting global indices.\\
 
 For clarity, the following description concentrates on the construction of the global indices for the sector-level of the classification. The methods used for constructing global indices for other classification levels and the single-level classification follow a similar approach.
 Let $d$ be the number of sectors and $\ell$ be the number of countries. Thus, we have $d=11$ sectors and $\ell = 67$ countries for the data we utilise.
 Let $S^{j,k}(t)$ denote the index value in US Dollars at time $t$ in respect of sector $j$ and country $k$, where $j\in\{1,2,3,\ldots ,d\}$
 and $k\in\{1,2,3,\ldots ,\ell \}$.
 Let $\bar\kappa$ denote the {\em proportional transaction cost} per US Dollar of turnover, where we fix, for simplicity, $\bar\kappa = 0.004$, which means $40$ basis points (bp) transaction cost. We calculate the $EWId(t_i)$, with initial value $EWId(t_0)=1$,  using the following recursive formulae: 
 $$
 EWId(t_i^{-}) = EWId(t_{i-1}) \times \sum_{j=1}^d\sum_{k=1}^\ell  \pi^{j,k}(t_{i-1}) \frac{S^{j,k}(t_i)}{S^{j,k}(t_{i-1})} ,
 $$
 $$ TO(t_i)=
 \sum_{j=1}^d \sum_{k=1}^\ell 
 \left|
 \pi^{j,k}(t_{i}) EWId(t_{i}^{-})
 -
 \pi^{j,k}(t_{i-1}) EWId(t_{i-1}) \frac{S^{j,k}(t_i)}{S^{j,k}(t_{i-1})} \right|, 
 $$
 and 
 $$
 EWId(t_i) = EWId(t_i^{-}) - \bar\kappa \times TO(t_i). 
 $$
 The weights in the above formulae are calculated using the formula
 
 $$
 \pi^{j,k}(t_i) = 
 \begin{cases} 
 	1/\{ N_\text{sector}(t_i)\, N^{j}_\text{sector, country}(t_i)\}&,\text{ if $S^{j,k} (t_i)>0$;}\\
 	0&,\text{ otherwise},
 \end{cases}
 $$
 where
 \begin{equation*}
 	N_\text{sector}(t_i) = \# \{ j\in\{1,2,\ldots ,d\} |   S^{j,k}(t_i)>0 \text{ for some } k\in\{ 1,2,\ldots , \ell \}  \} \end{equation*}
 denotes  the number of sectors that are present in any country at the time $t_i$, and
 \begin{equation*}
 	N^{j}_\text{sector, country}(t_i) = \# \{ k\in\{1,2,\ldots ,\ell\} |   S^{j,k}(t_i)>0  \} 
 \end{equation*}
 denotes the number of countries that have stocks in the $j$-th sector at this time.
 We start by equally weighting the market capitalisation-weighted sector indices from those countries with stocks in the sector to create an equally weighted world-sector index. Then, we equally weight all  these world sector indices to obtain the final global sector level index, the EWI11.\\
 \subsection{Comparison of EWIs and MSCI}
\begin{figure}
 	\includegraphics[width=\textwidth]{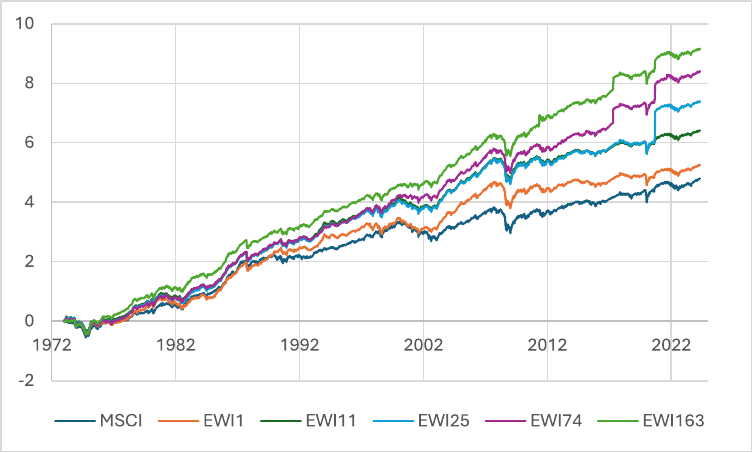}
 	\caption{Logarithm of the MSCI, and the EWId${}$, for  $d\in\{ 1,11,25,74,163\}$.}
 	\label{Fig2}
 \end{figure}

 In Figure~\ref{Fig2}, the logarithms of the EWId${}$ for  $d\in\{ 1,11,25,74,163\}$, and  the total return index of the world stock market, the MSCI, are plotted, where we scale all portfolios to $1$ at the initial observation time. 
 It is visually apparent that all EWIs exhibit higher average 
 growth rates than that of the MSCI over the period from 1973 until 2024. 
 The EWI163 is, due to its highest average growth rate, the best match of the benchmark. As shown in Theorem 3.6 in \citeN{Platen26e},  when all independent sources of randomness are represented by independently evolving stock portfolios, the factors, the GOP results from assigning equal weights to these portfolios. The EWI163 provides with its constituents the best approximation of factors in the set of EWIs.
 
 In practice, forming truly independent stock portfolios is challenging. Still the above construction of  EWIds  moves,  for increasing number $d$, in this direction. The logarithms of the EWIds, presented in Figure \ref{Fig2}, demonstrate that incorporating increasingly diverse sources of randomness through  distinct stock portfolios by applying equal weighting, results in higher average  growth rates. Consequently, this approach yields progressively improved proxies for the benchmark. The EWI163 is theoretically expected to form the best proxy of the benchmark and    to outperform significantly the average  growth rate of the MSCI. These predictions are confirmed  by  empirical results for the estimated log-return densities shown below. \\

 \section{Log-Return Densities}\label{sec:SGH}
 
 \subsection{Characterisation of Log-Return Densities}\label{logreturn}
 
 We have  matched almost all ten empirical facts, listed in the introduction,  with those properties predicted by the proposed model. It remains to check the listed empirical fact (2.), which states that estimated log-return densities for  proxies of the benchmark closely resemble  Student $t$ densities with slightly less than four degrees of freedom. The current paper tests below whether  this property cannot be easily rejected for the constructed set of daily observed EWIds and the MSCI. \\

 We capture the  daily observed log-returns $x_1,x_2,\ldots, x_i$ of the considered stock portfolios, namely, the MSCI, the EWI1, 
 the EWI11, the EWI25, the EWI74, and the EWI163 using the symmetric generalised hyperbolic (SGH) density, which is given in \citeN{Barndorff77}. Our study employs this rather general density 
  to demonstrate the  fit of the nested Student $t$ density to log-returns of various stock indices.
 The SGH density is described by the density of the random variable $X$ defined via
 $$X = \mu + \sqrt{V} \, Z,$$ 
 where $V$ has a generalised inverse Gaussian (GIG) density, i.e., $V \sim GIG(\alpha,\delta,\lambda  )$, and $Z\sim N(0,1)$.
 The GIG density of $V$ equals
 \begin{equation}\label{Eqn:fV}
 	f_V(v) = \frac{\alpha^\lambda }{2\delta^\lambda  K_\lambda (\alpha\delta )} v^{\lambda -1} \exp\left\{ -\frac{1}{2} (\alpha^2 v + \delta^2/v) \right\} 
 \end{equation}
 for $v>0$.
 Here, $K$ is the modified Bessel function of the second kind with index $\lambda$, given by the integral
 $$
 K_\lambda (\omega) = \frac{1}{2}\int_0^\infty t^{\lambda - 1} \exp \left(-\frac{1}{2}\omega (t+1/t)\right)\, dt .
 $$
 The $k$-th moment of $V$ equals
 \begin{align}
 	E[V^k]
 	&= \int_0^\infty v^k  \, f_V(v)\, dv \\ \notag
 	&= \frac{\alpha^\lambda }{2\delta^\lambda  K_\lambda (\alpha\delta )} \int_0^\infty v^{k+\lambda -1} \exp\left( -\frac{1}{2} (\alpha^2 v + \delta^2/v) \right) \, dv \\ \notag
 	&= \frac{\alpha^\lambda }{2\delta^\lambda  K_\lambda (\alpha\delta )}
 	\times 
 	\frac{2\delta^{k+\lambda}  K_{k+\lambda} (\alpha\delta )}{\alpha^{k+\lambda} } \\ \notag 
 	&= \frac{\delta^{k}  K_{k+\lambda} (\alpha\delta )}{\alpha^k  K_\lambda (\alpha\delta )}
 \end{align}
 and, particularly, one has
 $$
 E[V] = \frac{\delta  K_{\lambda + 1} (\alpha\delta )}{\alpha  K_\lambda (\alpha\delta )} .
 $$
 Furthermore, $X$ has the normal variance-mixture density 
 \begin{align}
 	\label{Eqn:fX}
 	f_X(x) &= \int_0^\infty \frac{1}{\sqrt{2\pi v}} \exp \left( -\frac{(x-g )^2}{2v}  \right) \, f_V(v)\, dv \\ \notag
 	&=\frac{1}{\delta K_\lambda (\bar{\alpha})} \sqrt{\frac{\bar{\alpha}}{2\pi }} \left( 1 + \frac{(x-g )^2}{\delta^2} \right)^{(\lambda-\frac{1}{2})/2} K_{\lambda - \frac{1}{2}} \left( \bar{\alpha} \sqrt{1 + \frac{(x-g )^2}{\delta^2}} \right) ,
 \end{align}
 where $x \in {\bf
 {R}}$. Here, $g\in {\bf
 {R}}$ is the location parameter and by defining $\bar{\alpha} = \alpha\delta $  we have that $\bar{\alpha}$ and  $\lambda$ account for the shape of the density. For interpretability, we define the scale parameter $\sigma$ as
 $$\sigma = 
 \begin{cases} 
 	\sqrt{2\lambda / \alpha^2}  &,\, \mathrm{if}\, \delta = 0, \,\mathrm{for}\, \lambda>0,\, \bar{\alpha}=0,\\ 
 	\frac{\delta^2 K_{\lambda+1}(\bar{\alpha})}{\bar{\alpha}\, K_\lambda (\bar{\alpha} )}	&,\, \mathrm{otherwise},
 \end{cases}
 $$
 which ensures that the expected value of the variance distribution, whose density is given in \eqref{Eqn:fV}, equals $\sigma$. Hence, the SGH density has the four parameters $g$, $\sigma$, $\bar{\alpha}$ and $\lambda$.
 
 The SGH density covers, as particular cases, the variance-gamma (VG), hyperbolic (H), normal-inverse Gaussian (NIG), Student $t$ (T), and normal (N) densities. These densities are nested in the SGH density and summarised in Table~\ref{Tab:SGHCases} for the respective parameter combinations.
 
 \begin{table}
 	\begin{tabular}{|l|l|}
 		\hline
 		Distribution &  Conditions on SGH Parameters\\ \hline
 		Variance Gamma &  $\alpha = \sqrt{2\lambda}/\sigma$, $\delta = 0$, $\lambda>0$\\
 		Hyperbolic &  $\lambda = 1$\\
 		Normal-Inverse Gaussian &  $\lambda = -1/2$\\
 		Student $t$ &  $\alpha = 0$, $\delta = \sigma \sqrt{\nu - 2}$, $\lambda = -\nu/2 <0$\\
 		Normal &  $\alpha = 0$, $\delta = \sigma \sqrt{\nu - 2}$, $\lambda = -\nu/2 <0$, $\nu\to\infty$ \\
 		\hline
 	\end{tabular}
 	\caption{\label{Tab:SGHCases}Table showing particular cases of the SGH density.}
 \end{table}
 
 \subsection{Estimating Log-Return Densities}\label{sec:fitting}

 We employ maximum likelihood estimation of the parameters when fitting the above described densities.
 The SGH density is the  nesting density, with the fitted parameters shown in Table~\ref{Tab:SGH}. The near-zero values of $\bar{\alpha}$ combined with the negative values of $\lambda$ indicate that a Student $t$ density is most likely the best fit to EWId and MSCI log-returns.\\
 
 \begin{table}\centering
 	\begin{tabular}{|l|l|l|l|l|l|} \hline
 		Index & $g$ & $\sigma$ & $\bar{\alpha}$ & $\lambda$ & log-likelihood \\ \hline
 		MSCI & 0.000357 & 0.008229 & 0.096962 & -1.719 & 46994.43 \\
 		EWI1 & 0.000391 & 0.006937 & 0.202617 & -1.457 & 49524.41 \\
 		EWI11 & 0.000477 & 0.006647 & 0.132465 & -1.555 & 50138.37 \\
 		EWI25 & 0.000550 & 0.007083 & 0.000010 & -1.580 & 49698.10 \\
 		EWI74 & 0.000625 & 0.007022 & 0.000055 & -1.599 & 49733.07 \\
 		EWI163 & 0.000681 & 0.007078 & 0.000030 & -1.622 & 49519.80 \\
 		\hline
 	\end{tabular}
 	\caption{\label{Tab:SGH}Fitted parameter values and log-likelihood values of the SGH density.}
 \end{table}
 
 In the next step, we fit  special cases of the SGH density.
 The variance-gamma density and its application to modeling stock market returns is described in \citeN{MadanSe90}. The results of fitting this density to  log-returns are shown in Table~\ref{Tab:VG}, where it is evident that the log-likelihoods are lower than those of the SGH density shown in Table~\ref{Tab:SGH}. In particular, for the EWI25, EWI74 and EWI163, the fits of the variance-gamma density are poor.
 
 \begin{table}\centering
 	\begin{tabular}{|l|l|l|l|l|l|} \hline
 		Index & $g$ & $\sigma$ & $\lambda$ & log-likelihood \\ \hline
 		MSCI & 0.000357 & 0.007836 & 1.17905 & 46871.35 \\
 		EWI1 & 0.000391 & 0.006566 & 1.04051 & 49408.71 \\
 		EWI11 & 0.000477 & 0.006260 & 1.07865 & 50002.64 \\
 		EWI25 & 0.000550 & 0.006617 & 0.98778 & 49385.02 \\
 		EWI74 & 0.000625 & 0.006600 & 1.00071 & 49399.51 \\
 		EWI163 & 0.000681 & 0.006668 & 1.04051 & 49203.09 \\
 		\hline
 	\end{tabular}
 	\caption{\label{Tab:VG}Fitted parameter values and log-likelihoods of the variance-gamma density.}
 \end{table}
 
 The hyperbolic density is proposed in  models by \citeN{EberleinKe95} and \citeN{KuchlerNeSoSt99}. Applying this density to our data, the estimated parameters and log-likelihoods are shown in Table~\ref{Tab:Hyperbolic}, where it is apparent that the fits are superior to those of the variance-gamma  density  but  inferior to those of the SGH density.
 
 \begin{table}\centering
 	\begin{tabular}{|l|l|l|l|l|l|} \hline
 		Index & $g$ & $\sigma$ & $\bar{\alpha}$ & log-likelihood \\ \hline
 		MSCI & 0.000357 & 0.007804 & 0.375202 & 46890.07 \\
 		EWI1 & 0.000391 & 0.006526 & 0.207777 & 49417.84 \\
 		EWI11 & 0.000477 & 0.006222 & 0.260464 & 50017.81 \\
 		EWI25 & 0.000550 & 0.006538 & 0.201636 & 49397.99 \\
 		EWI74 & 0.000625 & 0.006514 & 0.235913 & 49414.39 \\
 		EWI163 & 0.000681 & 0.006595 & 0.276294 & 49222.94\\
 		\hline
 	\end{tabular}
 	\caption{\label{Tab:Hyperbolic}Fitted parameter values and log-likelihood values of the hyperbolic density.}
 \end{table}
 
 The normal-inverse Gaussian (NIG) density is investigated in \citeN{Barndorff97}, and fitting this density to our  log-returns gives the results in Table~\ref{Tab:NormalInverseGaussian}. The log-likelihoods of the fitted NIG densities improve upon those of the hyperbolic and variance-gamma densities, but fall short of those of the  SGH density. By allowing $\lambda$ to be negative, we obtain improvements upon the hyperbolic and variance-gamma densities.
 
 \begin{table}\centering
 	\begin{tabular}{|l|l|l|l|l|l|} \hline
 		Index & $g$ & $\sigma$ & $\bar{\alpha}$ & log-likelihood \\ \hline
 		MSCI & 0.000357 & 0.007994 & 0.662131 & 46963.78 \\
 		EWI1 & 0.000391 & 0.006739 & 0.547556 & 49502.98 \\
 		EWI11 & 0.000477 & 0.006424 & 0.566493 & 50109.33 \\
 		EWI25 & 0.000550 & 0.006846 & 0.476975 & 49587.98 \\
 		EWI74 & 0.000625 & 0.006814 & 0.492485 & 49612.27 \\
 		EWI163 & 0.000681 & 0.006871 & 0.522416 & 49406.50\\
 		\hline
 	\end{tabular}
 	\caption{\label{Tab:NormalInverseGaussian}Fitted parameter values and log-likelihood values of the normal-inverse Gaussian density.}
 \end{table}
 
 The Student $t$ density is proposed in \citeN{Praetz72} for modeling weekly observations of 
 share-price indices from the Sydney Stock Exchange. When applied to fitting our log-returns, the results are shown in Table~\ref{Tab:Studentt}, where it is apparent that the log-likelihoods are, in respect of the EWI25, EWI74, EWI163 indicies, to three decimal places the same as those of the SGH density given in Table~\ref{Tab:SGH}. 
 \begin{table}\centering
 	\begin{tabular}{|l|l|l|l|l|l|} \hline
 		Index & $g$ & $\sigma$ & $\nu$ & log-likelihood \\ \hline
 		MSCI & 0.000357 & 0.005418 & 3.480912 & 46994.36 \\
 		EWI1 & 0.000391 & 0.004344 & 3.152811 & 49523.01 \\
 		EWI11 & 0.000477 & 0.004179 & 3.216502 & 50137.88 \\
 		EWI25 & 0.000550 & 0.004292 & 3.159123 & 49698.10 \\
 		EWI74 & 0.000625 & 0.004297 & 3.197261 & 49733.07 \\
 		EWI163 & 0.000681 & 0.004390 & 3.245581 & 49519.80 \\
 		\hline
 	\end{tabular}
 	\caption{\label{Tab:Studentt}Fitted parameter values and log-likelihood values of the Student $t$ density.}
 \end{table}
 Finally, the results for the normal density are shown in Table~\ref{Tab:Normal}, where the log-likelihoods are  worse than those of the other densities.\\
 
 \begin{table}\centering
 	\begin{tabular}{|l|l|l|l|l|l|} \hline
 		Index & $g$ & $\sigma$ & log-likelihood \\ \hline
 		MSCI & 0.000357 & 0.008229 & 45459.53 \\
 		EWI1 & 0.000391 & 0.006930 & 47769.76 \\
 		EWI11 & 0.000477 & 0.006630 & 48371.10 \\
 		EWI25 & 0.000550 & 0.010296 & 42447.33 \\
 		EWI74 & 0.000625 & 0.008846 & 44486.86 \\
 		EWI163 & 0.000681 & 0.008071 & 45720.54 \\
 		\hline
 	\end{tabular}
 	\caption{\label{Tab:Normal}Fitted parameter values and log-likelihood values of the normal density.}
 \end{table}
 \begin{table}
 	\small
 	\centering
 	\begin{tabular}{|l|l|l|l|l|l|l|l|} \hline
 		& & & & & & &\\
 		Index & $g$ p.a. & $\sigma$ p.a. & $\bar{\alpha}$ & $\lambda$ & $\nu = - 2 \lambda$ & log- & Best\\ 
 		& & & & & & likelihood & model\\
 		& & & & & & &\\	
 		\hline
 		& & & & & & &\\
 		MSCI & 0.092883 & 0.132690 & 0.096962 & -1.719 & 3.44 & 46994.43 & SGH\\
 		EWI1 & 0.101776 & 0.111741 & 0.202617 & -1.457 & 2.91 & 49524.41 & SGH\\
 		EWI11 & 0.124043 & 0.106853 & 0.132465 & -1.555 & 3.11 & 50138.37 & SGH\\
 		EWI25 & 0.142880 & 0.166010 & 0.000010 & -1.580 & 3.16 & 49698.10 & SGH\\
 		EWI74 & 0.162391 & 0.142644 & 0.000055 & -1.599 & 3.20 & 49733.07 & SGH\\
 		EWI163 & 0.177133 & 0.130138 & 0.000000 & -1.623 & 3.25 & 49519.80 & Student $t$\\ 
 		& & & & & & &\\
 		\hline
 	\end{tabular}
 	\caption{\label{Tab:Parameters} Parameter values of best fitting log-return densities.}
 \end{table}
 The best fitting models are deduced from Tables~\ref{Tab:SGH}, \ref{Tab:VG}, \ref{Tab:Hyperbolic}, \ref{Tab:NormalInverseGaussian}, \ref{Tab:Studentt} and \ref{Tab:Normal}, and are shown in Table~\ref{Tab:Parameters}.
 It is evident from the table that annualised average growth rates of over $0.17$ and annualised volatilities less than $0.11$ are attainable for the indices EWI11, EWI25, EWI74 and EWI163. This contrasts with the annualised average growth rate of $0.09$ and annualised volatility of $0.13$ of the MSCI. Note that the EWI163 has approximately the same volatility as the MSCI but an average annualised growth rate that is almost $9$\% higher. Since $\bar \alpha$ is for the EWI163 in Table \ref{Tab:Parameters} up to six decimals shown to be zero, the Student $t$ density  for the log-returns of the benchmark, predicted in \citeN{Platen26e}, manifests clearly. The best theoretical model fit is achieved by the EWI163, which identifies the predicted Student $t$ density as the log-return density in the rather large class of SGH densities. All other EWIds identify SGH densities with  $\bar \alpha$ closer to zero for increased $d$, which identifies SGH densities that are very close to a respective Student $t$ density. The degrees of freedom $\nu=-2\lambda$ are in all cases smaller than $4$ and larger than $2.9$. Closest to $4$ come  the degrees of freedom for the MSCI, which has a trajectory that shows no jumps besides the one at the 1987 Stock Market Crash, where some markets were closed for several days. \\One notices in Figure \ref{Fig2}   significant upward jumps in 2020 in the trajectories of the EWI25, EWI74, and EWI163. These jumps are caused by a single significant price jump of a Venezuelan stock. If we would remove the corresponding log-return from the data set, then the estimated degrees of freedom would move closer to four. \\

 We notice in Table \ref{Tab:Parameters} that the degrees of freedom for the EWIds are smaller than those of the MSCI and slightly lower than  four. The current paper explains this fact  by the randomness of the activity.  The activity times of the considered stock portfolios have    random derivatives. These generate in the squared volatility \eqref{squvola} some extra randomness, which means that there is some additional mixing of the normal variance-mixing log-return densities. 
  The   randomness of the activity in calendar time and occasional jumps make the estimated log-returns in calendar time  more leptokurtic, which leads  to lower degrees of freedom in the class of Student $t$ densities.

 \subsection{Likelihood Ratio Test}
 For testing the estimated log-return densities, we employ the maximum likelihood method, which relies on the predicted probability density as well as the empirical probability density generated by the data.
 Since the Student $t$ and the other considered densities are nested within the SGH density, a {\em likelihood ratio test}, as described in \citeN{NeymanPe28}, can be employed to compare these densities. We use the maximum likelihood method to estimate parameters for the observed log-returns and apply the  likelihood ratio test to assess which density is fitting best the data.

 We define the {\em likelihood ratio\/} in the form
 \begin{equation}
 	\label{a.2.55} \Lambda = \frac{\mathcal{L}^*_{model}}{\mathcal{L}^*_{nesting\ model}}.
 \end{equation}
 Here $\mathcal{L}^*_{model}$ represents the maximised likelihood function
 of a hypothesised model density, say, with $q$ parameters. On the
 other hand, $\mathcal{L}^*_{nesting\ model}$ denotes the maximised likelihood
 function for the density of a more general model, the nesting model, which has, say
 $q+p$ parameters and nests the hypothesised model density, for $q,p
 \in \{1,2,...\}$.
  It has been shown in \citeN{Rao73}
 that the density of the {\em test statistic}
 \begin{equation}
 	\label{a.2.57}
 	L = -2\,\ln (\Lambda)
 \end{equation}
 is asymptotically a chi-squared density for increasing number  of
 observations. Here the number of degrees of freedom $p$
 equals the difference between the number of parameters in the
 nesting model density and the hypothesised model density.\\
 \begin{table}[tb]
 	\centering
 	{\small
 	 		\begin{tabular}{|c||c|}
 			 \hline \hspace*{0.5cm}
 			$\chi^2_{0.99,1}$ \hspace*{0.5cm} & \hspace*{0.5cm} 6.635
 			\hspace*{0.5cm}  \\ \hline $\chi^2_{0.95,1}$
 			& 3.841  \\ \hline $\chi^2_{0.90,1}$ & 2.706  \\ \hline
  			$\chi^2_{0.20,1}$ & 0.064  \\ \hline
 			$\chi^2_{0.10,1}$ & 0.0158  \\ \hline
 			$\chi^2_{0.05,1}$ & 0.0039  \\ \hline
 			$\chi^2_{0.01,1}$ & 0.000157  \\ \hline
 			$\chi^2_{0.001,1}$ & 0.000002  \\ \hline
 	\end{tabular}} 
 	\caption{Quantiles for the chi-squared distribution with one degree of freedom.}	\label{tab:2.3}
  \end{table}
 
 Specifically, the SGH density is a four-parameter density, while the four special cases we consider, the  variance-gamma, Student $t$, hyperbolic, and normal-inverse Gaussian densities, are three-parameter densities. This implies that their test statistic is chi-squared distributed with one degree of freedom. It can
 be shown, as detailed in \citeN{Rao73}, that as the number of observations tends to infinity, 
 we have
 \begin{equation}
 	\label{a.2.57_1}
 	P\left( L < \chi^2_{1-\alpha,1}\right) \approx
 	F_{\chi^2(1)} \left(\chi^2_{1-\alpha,1}\right) = 1-\alpha,
 \end{equation}
 where $F_{\chi^2(1)}$ denotes the chi-squared distribution with
 one degree of freedom and $\chi^2_{1-\alpha,1}$  its
 $100(1-\alpha)\%$ quantile. We list respective quantiles for one degree of freedom in Table \ref{tab:2.3}. \\

 This allows us to  check for a given
 $100\alpha \%$ confidence level whether the test statistic
 $L$ is in the $100 (1-\alpha)\%$ quantile of the chi-squared
 distribution with one degree of freedom. If the relation
 \begin{equation}
 	\label{a.2.57_2}
 	L < \chi^2_{1-\alpha,1}
 \end{equation}
 is not satisfied, then we  reject at the $100 \alpha \%$
 significance level the hypothesis that the suggested model is the
 true underlying model. Otherwise, we cannot falsify this hypothesis on the
 basis of this likelihood ratio test.\\
 \subsection{Log-Return Density}
 Recall that we estimate the densities of log-returns for fixed step size in calendar time. We do not remove in the construction of our  proxies of the benchmark any extreme log-returns as potential outliers from our data set. Therefore, market crashes and sudden market corrections are not discarded. The proper modeling of extreme log-returns is of great importance in risk management and removing outliers would not be appropriate. When testing daily log-returns of the considered stock portfolios, in every case, we found that either the Student $t$ density or the SGH density with small $\bar \alpha$ and degrees of freedom between $3$ and $4$ offer the best fit. \begin{table}
 	\small
 	\centering
 	\begin{tabular}{|l|l|} \hline
  		Index & $L = -2\,\ln (\Lambda)$  		 \\
 	 		\hline
 	 		MSCI & 0.00000298 
 		\\
 		EWI1 & 0.0000566 
 		\\
 		EWI11 & 0.0000194 
 		\\
 		EWI25 & 0.0000000107 
 		\\
 		EWI74 & 0.00000000771 
 		\\
 		EWI163 & 0.00000000000 
 		\\ 
 		\hline
 	\end{tabular}
 	\caption{\label{Tab:Ratiotest} Likelihood ratios of fitted Student $t$  log-return densities.}
 \end{table}
 
 In Table \ref{Tab:Ratiotest} we show in the second column the test statistics that emerge for the different nested models. 
 It follows that none of the test statistics is greater than $6\times 10^{-5}$. This means by Table \ref{tab:2.3} that
 the hypothesis that a Student $t$ density with between three and four degrees of freedom is the true log-return density of the considered indices cannot be rejected at the high    level of significance of $1$\%. For the EWI163 the level of significance is considerably lower. This is consistent with  the theoretical predictions of the proposed model for the benchmark dynamics. The likelihood ratio for the EWI163 in Table \ref{Tab:Ratiotest} is exceptionally close to zero. Thus,  all ten  empirical facts, listed in the introduction, are empirically shown for the considered proxies of the benchmark. In the sense, as formulated in the introduction, the proposed model is a realistic model for the benchmark, where the EWI163 turns out to represent the best proxy of the benchmark. This model can be employed for benchmark-neutral pricing and hedging, which yields for extreme-maturity zero-coupon bonds and many other derivatives significantly lower prices than risk-neutral pricing suggests, as described in \citeN{Platen25a}. \\

\section*{Conclusion}
The paper constructed a proxy for the  growth optimal portfolio (GOP) of the global stock market and derives  a realistic model for its dynamics. This model is based on four information-theoretic principles, which predict properties that match  a list of empirical facts, where the predicted log-return density is tested.  The normalized GOP is modeled as a Cox-Ingersoll-Ross (CIR) processes of dimension four that evolves in  activity time. The derivative of the normalized activity time follows in normalized activity time a CIR process of dimension one. The derived model offers explanations for various empirical facts, allows accurate benchmark-neutral pricing and hedging, and encourages econometric modeling of exchange rates and stocks. \\

 \section*{Data}
An EXCEL sheet is separately available, which covers  simulations and  calculations of the figures.  
\section*{Disclosure of Interest}
There are no interests to declare. 
\section*{Funding}
No funding was received.
\pagestyle{plain}
\bibliographystyle{chicago}
\bibliography{my}

\end{document}